# Agricultural On-Demand Networks for 6G enabled by THz Communication

Daniel Lindenschmitt[1,†], Christoph Fischer[2,*], Simon Haussmann[3,+], Marc Kalter[4,†], Ingmar Kallfass[5,+], and Hans D. Schotten[5,*†]

[†]Institute for Wireless Communication and Navigation. RPTU Kaiserslautern-Landau. D-67663 Kaiserslautern.
[*]German Research Center for Artificial Intelligence GmbH, DFKI. D-67663 Kaiserslautern.
[+]Institute of Robust Power Semiconductor Systems. University of Stuttgart. D-70569 Stuttgart.

[1]daniel.lindenschmitt@rptu.de, [2]christoph.fischer@dfki.de, [3]simon.haussmann@ilh.uni-stuttgart.de,
[4]marc.kalter@rptu.de, [5]ingmar.kallfass@ilh.uni-stuttgart.de, [6]hans_dieter.schotten@dfki.de, schotten@rptu.de

## Abstract

The transforming process in the scope of agriculture towards Smart Agriculture is an essential step to fulfill growing demands in respect to nourishment. Crucial challenges include establishing robust wireless communication in rural areas, enabling collaboration among agricultural machines, and integrating artificial intelligence into farming practices. Addressing these challenges necessitates a consistent communication system, with wireless communication emerging as a key enabler. Cellular technologies, as 5G and its successor 6G, can offer a comprehensive solution here. Leveraging technologies following the ITU-R M. 2160 recommendation like THz communication, low-latency wireless AI, and embedded sensing, can provide a flexible and energy-efficient infrastructure. This paper introduces on-demand networks based on the OpenRAN approach and a 7.2 functional split. By implementing THz front-hauling between components, a flexible application of 5G or future 6G networks can be realized. Experiments demonstrate that THz communication is suitable for data transmission over the eCPRI interface, particularly in terms of data rate, thereby reducing the need for wired alternatives such as fiber optic cables. Furthermore, limitations such as limited range are discussed, and possible initial solutions are presented. The integration of the OpenRAN standard further enhances flexibility, which is crucial in dynamic agricultural environments. This research contributes to the ongoing discourse on the transformative potential of 6G-enabled wireless communication in shaping the future of smart agriculture.

## 1 Introduction

The agricultural sector is a key topic for global consideration as it has a direct or indirect impact on human society as a whole. By meeting the demand for food, agriculture plays a fundamental role in shaping societal well-being and development. It is therefore essential that agricultural techniques meet both current and future demands due to growing population dynamics, ideally through environmentally sustainable methods.

Currently, agricultural production is insufficient to meet the increasing demand for food, and projections indicate that an increase in production of at least 60% will be required by 2050 [1]. Given the finite nature of arable land, an increase in agricultural efficiency is essential. One way to achieve this increase is through digitalization, which offers opportunities to optimize existing methods and the use of resources. A transforming process from conventional agriculture towards so called Smart Agriculture needs to address a broad field of challenges:

- How can we integrate a robust wireless communication system between all components of a system especially in the scope of rural areas with comparatively poor coverage or difficult terrain?
- How can an interaction and collaboration between agriculture machines be established and reliably be executed? • How can suitable artificial intelligence be included into the system to enhances processes for example in farming scenarios?

These challenges highlight the critical importance of implementing a unified communication system in the transformation process. In the context of agricultural scenarios, wired solutions are seen as insufficient to meet the evolving demand for adaptable connectivity, especially with regard to seasonal fluctuations in coverage requirements. As a result, wireless communication is proving to be a decisive factor in the realization of future agricultural concepts on the way to Smart Agriculture. Common non-cellular wireless communication technologies like Wi-Fi or LoRaWAN might be able to fulfill some of the rising demands; however, none of them can cover all aspects needed on the track to a successful transformation into Smart Agriculture [2]. The concept for on-demand networks presented in this paper is based on a new interpretation of Open - Radio Access Network (OpenRAN) functional splits with the inclusion of a THz wireless data link and thereby combines relevant aspects from current 6G research. THz bands are capable of providing large data rates, which have already been verified experimentally multiple times under outdoor conditions (see Section 5). Data rates over 100Gbps and distances of more than 500m are achieved. In addition, the system is built according to the OpenRAN standard and exploit its advantages, such as the use of general-purpose hardware to enable an even more cost-efficient and flexible application. This is particularly useful in scenarios that are subject to regular changes in coverage, purpose, and duration of use or cyclical repetitions, which is especially the case in the field of agriculture, but also with restrictions, for example, in the field of construction sites. Starting with Section 2, we introduce recent research on use cases of Terahertz communication in the scope of agriculture and then bring them in Section 3 into a context of 6G by defining requirements, key technologies, and relevant applications. After that, Section 4 examines the potential benefits of implementing 6G research such as THz communication in agriculture and discusses potential limiting factors. In addition, technical options for using on-demand networks and an OpenRAN functional split are presented. Then, Section 5 presents first results of a THz link between the radio unit and distribution unit as an enabler for a distributed system based on function sharing, followed by Section 6 with a conclusion and outlook on future work in this area.

## 2 Related Work

Unlike other 6G technologies, Terahertz technology has found extensive application in agriculture for sensing purposes. Nonetheless, due to ongoing climate shifts, researchers strongly believe that its potential within the realm of plant science remains largely unexplored, especially concerning its capacity to monitor plant health and biological conditions to sustain crop productivity. An illustrative case in this research field is a system capable of gauging water content and detecting pesticide residues in leaves using Terahertz technology. The study subjects plant leaves to Terahertz analysis, investigating the impact of thickness and water content on transmission loss and attenuation across frequency ranges. This pursuit yields valuable insights into detecting pesticides in leaves through THz. Although the experimental laboratory setup produced reliable measurements of transmission loss for diverse plant species, realworld implementation is still pending [3]. Terahertz technology is harnessed for detecting plant health, enabling nanoscale monitoring of plant nutrients to facilitate early identification of agricultural diseases and malnutrition [4]. Additionally, it serves to sense climate alterations, concurrently performing ultra-wideband communication and atmospheric sensing by leveraging Terahertz signal absorption. Analysis of signal path loss and power spectral density helps determine concentrations of climateaffecting gases [5]. Another technique reliant on Terahertz radiation aids in detecting grain storage quality, significantly reducing stored grain damage due to Terahertz radiation's traits of low energy, high permeability, and high signal-to-noise ratio [6].

These instances underscore Terahertz technology's prevalent use in sensing applications rather than data transmission. However, the ambitions of 6G communication involve exploiting Terahertz frequencies for high-data-rate communication, essential for future applications like holographic communication, extensive bio/nano-Internet of Things (IoT), haptic communication, and tele-operated driving [4]. OpenRAN, introduced within the 5G mobile communication standard, seeks to imbue the traditional closed and proprietary Radio Access Network (RAN) technology with openness and intelligence. By offering flexibility, improved performance, and cost-efficiency in RAN deployment and operation, OpenRAN emerges as a credible and potentially indispensable alternative to conventional proprietary RAN solutions. It's architecture, particularly employing the 7.2 split option, leads to simplified communication interference between OpenRAN units [7].

## 3 Wireless Communication in Agriculture

Two factors are driving agriculture today: increasing production volumes while improving product quality. Due to this, it will become necessary to make conventional technologies in the field of agriculture more efficient and to integrate new possibilities e.g. apply pesticides precisely. In this transformation process from classical agriculture to Smart Agriculture, connectivity is a crucial aspect to enable collaborative tasks and exchange of information. Therefore, efficient and robust communication technologies play a decisive role in the mentioned transformation process. Cellular technologies are a promising solution here, already the current 5G standard offers significant advantages in terms of data rate, coverage, and deterministic data transmission compared to other technologies such as Wi-Fi or LoRaWAN [8]. Besides of this, 5G also introduced NonPublic Networkss (NPNs), which enables not only traditional Mobile Network Operators (MNOs), but all landowners to operate their own mobile network in certain frequency ranges. The Recommendation ITU-R M. 2160 can be seen as an important basis for standardization towards 6G, which includes ubiquitous connectivity, massive communication, and AI as new metrics [9]. Potential key technologies like Terahertz communication with low latency and high transmission rate, embedded sensing, or wireless AI combined with a flexible and energy-efficient infrastructure can be used to tackle the questions we raised earlier. While a THz link is utilized as a front-hauling solution, it can be used more flexibly on the one hand, and on the other hand, it saves the deployment of new wired data links such as fiber optic lines.

Enabling new possibilities in the field of Smart Agriculture, such as collaborative, autonomous agricultural machinery are currently under research [10]. However, extensive hardware deployment is often required for operation, which means that many use cases cannot be operated economically. In addition, current 5G technologies are designed for static coverage, which in agriculture means that owners sometimes have to permanently license large areas of land to operate a corresponding 5G radio system here. This increases costs and energy consumption and reduces the sustainability of digitization.

The approach presented in this paper can provide the necessary tools for smart agriculture. A dynamic option for on-demand operation of a radio network, which on the one hand reduces operator costs and on the other hand makes a contribution to sustainable information and communication technologies [11], [12]. In this context, the efficient integration of artificial intelligence is also an important aspect. AI plays an important role here not only in the actual agricultural applications but also in the commissioning of the on-demand network. Here, interference with 6G networks must be prevented and, if necessary, interaction between them must take place in an automated manner. Spectrum allocation of neighboring on-demand networks as well as trusted data exchange and temporary licensing are topics where AI can support on the network side and thus enable simpler operation [13]. The aspect of dynamic, flexible and cost-efficient implementation is underlined in the presented concept by the integration of the OpenRAN standard. This can be applied not only in the field of agriculture but also in other vertical industries with similar constraints, especially in the area of dynamic network coverage [14]. Front-hauling based on THz wireless communication will be the enabler for flexible, high-data-rate communication. Like in E-band (60-90GHz), where two 5GHz windows are allocated for point-to-point links, wireless connectivity is already used frequently. But with its absolute bandwidth of over 70GHz, H-band (220-325GHz) is capable of fulfilling the bandwidth demands of even more data-intensive applications, which is also shown by Equation 1. The formulation of new standards in the H-band, like IEEE 802.15.3d, shows how pressing technical solutions in this band are needed. For a successful transformation from classical agriculture to Smart Agriculture, it will be necessary to enable an efficient way of Edge-Computing or Over-the-Air Computing for a central post-processing of data from various sensor systems, including ultrasound sensors, multiple high-resolution cameras, or positioning sensors with the lowest possible latency and without compression loss. Hereby the transfer of IQ data between Radio Unit (RU) and Distributed Unit (DU) is necessary for the introduced on-demand networks and the selected OpenRAN functional split. This will increase the efficiency of data processing in terms of power consumption and removes the necessary intelligence from the place of use, which is particularly important in agriculture.

## 4    Concept of On-Demand Networks

The sporadic use of the infrastructure is an important aspect, which has an impact on the reasonably low cost of the infrastructure to be deployed and also a positive effect on the sustainable usage with respect to the communication network. The seasonal aspect of infrastructure use is not only applied to agriculture but can also be transferred to other areas such as public events.

For this reason, on the one hand, an on-demand solution that enables a pay-as-you-go characteristic, so the communication network will only be active when needed, and, on the other hand, a minimally invasive setup for the network provider that requires a minimal amount of wired-based connectivity for rural areas. A farmer should, therefore, be able to rent the communication resource for dedicated areas when planning the farming of a field. The provisioning onsite then takes place through a setup, where the goal is to bundle as much complexity as possible in a central location to make the central node usable for as many field nodes as possible. The field nodes are, therefore, reduced in complexity and are connected wirelessly to make the field nodes fast and cost-efficiently deployable and power-efficient. The architecture will optimize the utilization of the central instance due to the combination of sporadic usage by multiple field nodes and can decrease complexity and setup due to a wireless connection of the field nodes. The on-demand concept presented is based on OpenRAN by using the idea of a functional split and extended with an intervening THz front-haul link. The necessary hardware can be permanently attached to an existing and suitable infrastructure, such as a wind turbine, or installed in a corresponding vehicle or trailer with a radio mast that can change its location. The second case offers the highest possible flexibility in terms of on-demand networks and is therefore used as an example for this paper. However, significantly higher requirements are also placed here, such as the exact, most autonomous possible antenna alignment of the THz front-hauling. The limitations of the presented concept are further discussed in Section 5.

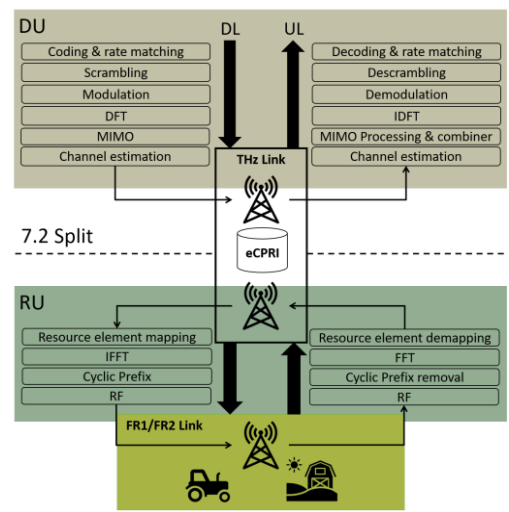

Figure 1 PHY-layer processing chain for Uplink (UL) and Downlink (DL)

## 4.1 OpenRAN

A fundamental principle of OpenRAN is to add standardized interfaces between these blocks to make them interchangeable and to be able to divide them into several building blocks that follow a fixed functional split. The exact distribution of the functions can be freely selected within the framework of various options. OpenRAN divides the RAN into three building blocks, as also standardized by 3rd Generation Partnership Project (3GPP) for 5G: the Centralized Unit (CU) for the higher layers of the RAN, the DU for MAC and higher PHY layers, and the RU for lower PHY operations. Due to the functional split, the functions in the CU and DU can be virtualized and run on commercial general-purpose hardware. Following OpenRAN, there are multiple options for a functional split that distributes functionalities between the three building blocks. For this paper, a 7.2 split was chosen, which is a common variant of the OpenRAN standard and shown in Figure 1.

The introduced concept of agriculture on-demand-networks will take advantage of the 7.2 functional split by placing CU and DU in a central node for efficient edge computing. The DU is the main processing unit and is responsible, among other things, for modulation, Discrete Fourier Transform (DFT), and generating the Multiple Input Multiple Output (MIMO) signals in the downlink (DL) or vice versa in the uplink (UL). Only the RU will stay at the field node, whereby the communication between the RU and DU is managed by the enhanced Common Public Radio Interface (eCPRI). The eCPRI is a standard interface used for connecting the mentioned components, providing a high-speed data link between the RU and DU, which is required to support the high bandwidth and low latency requirements of 5G [15]. Since many functions of the higher PHY are available in the DU, IQ data must be transferred to the RU. Equation 1 shows how to calculate the eCPRI IQ data rate used for the downlink [16]. The uplink might change depending on the quantization bit-width that is used here.

$$C_{FH} = \frac{B}{\Delta f} * \frac{1s}{t_{SF}} * N_{S/t_{SF}} * N_A * N_{AQ} * \psi_{DC} \quad (1)$$

Due to the high-speed data connection enabled by the eCPRI, it is necessary that the interface complies with strict restrictions in terms of end-to-end latencies with a maximum of $100\mu s$ and must also be able to process the corresponding bandwidth. Therefore, it is necessary to establish a correspondingly powerful connection between RU and DU, where so far no wireless solutions exists.

## 4.2 THz Front-Haul Link

By introducing a THz link between RU and DU and replacing the fiber connection, the restrictions of the eCPRI regarding latency and data rate can be fulfilled. Wireless frontor back-hauling is usually implemented as a point-to-point link. So, the physical layer and the data link layer have to provide full functionality, whereas the upper network layers just handle packets consecutively with no need for sharing between different nodes, sessions, or applications. Furthermore, reported THz experiments in Section 5 use a simple single-carrier modulation format without the challenges of real-time fourier transformation and multiple carrier synchronization efforts like in Orthogonal Frequency-Division Multiplexing (OFDM). This reduces the routing of information and computation, providing an overall low latency in the wireless link. Recent efforts in electronic mmW and sub-mmW transceiver design result in solutions for realtime communication in the sub-THz or THz domain and thus form the basis for the use of a front-haul link in the THz frequency for data transmission via the eCPRI interface.

State-of-the-art compound semiconductor technologies are pushing the limits of ultra-broadband circuits at 300GHz towards higher performance. Output powers well above 10dBm are reported by several foundries. Just some examples of leading technologies are: an 80nm InP MOS HEMT, a 35nm InGaAs mHEMT, or a 250nm InP HBT technology. These technologies are demonstrating high-power amplifiers in H-band with 16.8, 13.8, and 10.6dBm saturated output power with bandwidths of 19, 48, or 80GHz, respectively [17], [18], [19]. The center frequencies of these amplifiers are designed to 275, 304, and 260GHz.

## 5 Technical realization & Results

### 5.1 Setup

A harvesting scenario was selected for an exemplary realization. In this setup, the on-demand RU is located directly at the place of use, e.g. in the form of a trailer for spanning, a local cellular network in the area of FR1 or FR2 and additionally via a point-to-point front-haul connection in the THz range for transmitting the local IQ data via the eCPRI interface to the DU, e.g. located in a central data center. Figure 2 is illustrating the chosen setup.

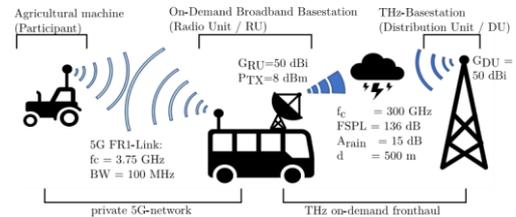

Figure 2 Overview of the on-demand THz front-hauling scenario

The DU can have one or more THz transceiver in order to connect to multiple RUs concurrently. To be noted is that the number of transceivers on the DU does not have to scale with the number of connected RUs but each transceiver should be able to connect various RUs, for example by automatic adjustable antennas. Even multiple RUs can share a single THz transceiver, which then restricts the usage of the connected RUs to the maximum throughput of the THz connection.

$$C_{exp} = \frac{100\ MHz}{15\ kHz} * \frac{1\ s}{1\ ms} * 14 * 8 * 16 * \frac{4}{3} \approx 16 Gpbs \quad (2)$$

The connection between RU and DU is the crucial aspect here, realized typically over fiber, which brings low overhead and high data rate optimized streaming protocol, as already introduced in Section 4. The parameters listed in table 1 are selected for the technical realization in FR1 as it might be used for private networks as in Germany for frequencies 3.7-3.8GHz and typical ones for 5G. This results in the number of subcarriers being approximately 6000, with 14 symbols per subframe and a subframe duration of 1ms. Based on the equation 1 and 2, the minimum required data rate for a wireless eCPRI front-haul link in the described scenario is around 16 Gbps. This data rate scales with parameters like bandwidth or number of antennas used, due to digital beamforming, and therefore is expected to be way higher in FR2. Another aspect worth mentioning is the timing restrictions that the RAN has on this interface, which is 100 $\mu$s end-to-end latency budget for IQ data between the RU and DU [20].

**Table 1** Exemplary parameter set to calculate front-haul capacity in downlink based on [16]

| Parameter | Value |
|---|---|
| Bandwidth $B$ | 100 MHz |
| Subcarrier spacing $\Delta f$ | 15 kHz |
| Subframe period $t_{SF}$ | 1 ms |
| OFDM Symbols per subframe $N_{S/t_{SF}}$ | 14 |
| Number of Antennas $N_A$ | 8 |
| Quantization bit-width per sample $N_{AQ}$ | 8*2=16 |
| Front-haul Overhead $\psi_{DC}$ | 4/3 |

## 5.2 Front-haul Experiments

The requirements for on-demand networks were determined in the previous section; the data rate of at least 16 Gpbs is particularly relevant for the evaluation of the present experiments. Table 2 provides an overview of recent communication experiments compared to existing standards. In most experiments, simplex data transmission is achieved, but also duplex transmissions are reported. For the agricultural use case, a heavily asymmetrical link, or even simplex data transmission, is sufficient because high data quantities are uploaded from the sensors to the DU, but just a few instructions have to be sent back to the agricultural machine. In these frequencies, purely electronic and opto-electronic solutions are present. Opto-electronic solutions often have the advantage of higher spectral purity and thus better signal quality. A downside is less available signal power.

Based on the results presented, it was shown that ondemand networks with THz front-haul link meet the requirements of OpenRAN and can therefore be used for a wireless connection between RU and DU. It was also shown that novel THz technologies can achieve ranges similar to previous 5G coverage under normal conditions. Potential limitations of the concept are discussed in the following subsection.

## 5.3 Limitations

The range of the THz link is a limiting factor in the concept. This issue could be addressed by establishing multihop links between more than one station. Again, the hard requirements of the eCPRI to meet timing restrictions may pose a problem for this link. Further concepts can be developed that are based on the fact that coordinated channel access can be specified by a deterministic number of clients, which in turn can have a positive effect on the timing restrictions of the eCPRI. For the targeted application, however, the availability of the link should be given under any outdoor weather condition. Foggy or heavy rainy conditions appearing in agricultural environments in many scenarios, like in early morning situations, have to be handled. Without clear line-of-sight, the usage of optical links is not possible.

For the Smart Agriculture use cases introduced in this paper, additional to the analog frontends, highly directive parabolic antennas with an antenna gain of around 50dBi are used to account for the high loss transmission of the static point-to-point link, which, as a consequence, results in the need for beam-steering and tracking. An overview is given in Figure 2. Although the RU is represented as a relocatable vehicle in the setup described in this work, the RU is stationary and does not move during use. Hence no continuous beam steering is needed. However, the concept introduced could also be supplemented more clearly by a mobile component in its fields of application and thus make an important contribution to flexible communication networks [27].

**Table 2** High Throughput Communication Experiments in H-Band

| Ref. | Technology | Center Frequency | Data rate | Distance | Signal generation | Architecture |
|---|---|---|---|---|---|---|
| [21] | 80 nm InP HEMT | 290GHz | 100Gbps | 54m | electronic | simplex |
| [22] | 35 nm InGaAs mHEMT | 300GHz | 12.5Gbps | 645m | opto-electronic | duplex |
| [23] | 35 nm InGaAs mHEMT | 300GHz | 32Gbps | 180m | electronic | simplex |
| [24] | 50 nm InGaAs mHEMT | 240GHz | 96Gbps | 40m | electronic | simplex |
| [25] | 50 nm InGaAs mHEMT | 300GHz | 102.4Gbps | 500m | opto-electronic | simplex |
| [26] | WiFi 6 (IEEE 802.11ax) | 6GHz | 9.6Gbps | 50m | electronic | duplex |
| [10] | 3GPP 5G Rel. 15 | 3.75GHz | <1Gbps | 500m | electronic | duplex |

With a wireless THz-path of 500m, the signal power at the receiver can be estimated to $P_{RX}=-48$dBm according to the link budget calculation. It considers an average transmit power of 8dBm, the Free Space Path Loss of 136dB, the antenna gains and an additional atmospheric, weather and packaging loss of 15dB. By considering an antenna temperature of 300K and a typical receiver noise figure of 5dB, the spectral noise density can be calculated to $S_{AWGN} = 1.29 \times 10^{-17}$ mW/Hz by utilizing the Friis formula and the Boltzmann constant. With these given values, the bandwidth for a chosen Carrier-to-Noise-Ratio (CNR) can be approximated with Equation 3.

$$BW = \frac{10^{(P_{RX}-CNR)/10} \times 1mW}{S_{AWGN}} \quad (3)$$

For a CNR of 20dB, a bandwidth of 12.3GHz can be chosen. In this calculation, other non-idealities of transceivers like spurious tones are not regarded, but the calculation shows the potential of the hardware. Depending on the scenario and distance, less directive antennas with 30dBi can be employed, decreasing the link budget, but a larger area can be covered with a larger beam width and less stringent beam-steering requirements. As a result, an easier deployment and installation of THz base stations would be possible, but consequently, less bandwidth, CNR, and/or distance would be achievable.

## 6 Conclusion

In this paper, we have presented a concept for agricultural on-demand networks using emerging 6G technologies, such as Terahertz communications, together with OpenRAN and outlined their benefits. First, previous work in THz communication in the field of agriculture was presented and discussed. Then, in Section 3, the need for a reliable and dynamic communication system was explained and placed in the context of future wireless communication. In Section 4, we introduced the concept of an on-demand network based on the OpenRAN architecture and the functional split with a front-haul communication based on THz communication between Radio Unit and Distribution Unit. It was shown that, the concept of on-demand networks is able to provide robust, high-rate data transmission, while at the same time, it is a dynamic and flexible solution. We then presented a realistic setup in section 5 and determined the necessary data rate for typical 5G requirements between RU and DU, as well as using these results to discussed possible implementation approaches for the THz link in detail, while finally addressing potential limitations of the concept. In future work, the aspect of the strict latency requirements of the eCPRI will be investigated in more detail. The reported THz transmission experiments up to this day show distances up to 650 m, which are not sufficient to cover multiple fields in agricultural use cases. So, for a reasonable 6G solution, multi-hop scenarios have to be investigated in the future in terms of deployment cost and impact on overall latency. Furthermore, the presented functional split also includes the possibility to operate a moving RU, which makes tracking the THz link between RU and DU necessary. This will also be considered in more detail in future work.

## Acknowledgment

The authors acknowledge the financial support by the German *Federal Ministry for Education and Research (BMBF)* within the project »Open6GHub« {16KISK-003K, -004 & -019}.